\documentclass[journal]{IEEEtran}

\ifCLASSINFOpdf
\else
   \usepackage[dvips]{graphicx}
\fi
\usepackage{url}
\usepackage{bm}
\usepackage{color}

\usepackage{graphicx}
\usepackage{booktabs}
\usepackage{makecell}
\usepackage{comment}
\usepackage{hyperref}
\usepackage{tikz}
\usepackage{textcomp}
\usepackage{lipsum}
\usepackage{setspace}

\begin{document}

\title{GAMIVAL: Video Quality Prediction on Mobile Cloud Gaming Content}

\author{Yu-Chih~Chen,
        Avinab~Saha,
        Chase~Davis,
        Bo~Qiu,
        Xiaoming~Wang,
        Rahul~Gowda,
        Ioannis~Katsavounidis,
        and~Alan~C.~Bovik,~\IEEEmembership{Fellow,~IEEE}
        
\thanks{The work of Alan C. Bovik was supported by the National Science Foundation AI Institute for Foundations of Machine Learning (IFML) under Grant 2019844. This work was supported by Meta Platforms Inc.}
\thanks{Yu-Chih~Chen, Avinab~Saha and Alan~C.~Bovik are with the Laboratory for Image and Video Engineering (LIVE), Department of Electrical and Computer Engineering, The University of Texas at Austin, Austin, TX, 78712 USA (email: berriechen@utexas.edu; avinab.saha@utexas.edu; bovik@ece.utexas.edu).}
\thanks{Chase~Davis, Bo~Qiu, Xiaoming~Wang, Rahul~Gowda and Ioannis~Katsavounidis are with Meta Platforms Inc., One Hacker Way Menlo Park, CA 94025 USA (email: chased@meta.com; qiub@meta.com; xmwang@meta.com; rahulgowda@meta.com; ikatsavounidis@meta.com).}}

\newcommand\copyrighttext{%
  \footnotesize \textcopyright Copyright © 2023 IEEE. Personal use of this material is permitted. However, permission to use this material for any other purposes must be obtained from the IEEE by sending a request to pubs-permissions@ieee.org. 
  DOI: \href{https://doi.org/10.1109/lsp.2023.3255011}{10.1109/LSP.2023.3255011}}
\newcommand\copyrightnotice{%
\begin{tikzpicture}[remember picture,overlay]
\node[anchor=south,yshift=10pt] at (current page.south) {\fbox{\parbox{\dimexpr\textwidth-\fboxsep-\fboxrule\relax}{\copyrighttext}}};
\end{tikzpicture}%
}

\maketitle
\copyrightnotice
\begin{spacing}{0.99}
\begin{abstract}
The mobile cloud gaming industry has been rapidly growing over the last decade. When streaming gaming videos are transmitted to customers' client devices from cloud servers, algorithms that can monitor distorted video quality without having any reference video available are desirable tools. However, creating No-Reference Video Quality Assessment (NR VQA) models that can accurately predict the quality of streaming gaming videos rendered by computer graphics engines is a challenging problem, since gaming content generally differs statistically from naturalistic videos, often lacks detail, and contains many smooth regions. Until recently, the problem has been further complicated by the lack of adequate subjective quality databases of mobile gaming content. We have created a new gaming-specific NR VQA model called the Gaming Video Quality Evaluator (GAMIVAL), which combines and leverages the advantages of spatial and temporal gaming distorted scene statistics models, a neural noise model, and deep semantic features. Using a support vector regression (SVR) as a regressor, GAMIVAL achieves superior performance on the new LIVE-Meta Mobile Cloud Gaming (LIVE-Meta MCG) video quality database.
\end{abstract}

\begin{IEEEkeywords}
image/video quality assessment, mobile cloud gaming, natural scene statistics, no-reference, perceptual quality, temporal statistics
\end{IEEEkeywords}

\IEEEpeerreviewmaketitle

\section{Introduction}

\IEEEPARstart{T}{he} development of broadband wireless Internet technology and mobile devices has significantly boosted the popularity of mobile games. Cloud gaming provides users a way to access many genres of games by remotely rendering streaming games in the cloud as videos. Client devices, such as smartphones and tablets capture users' interactions, and transmit them to cloud servers. This approach makes it possible to deliver high-computation video games to any modern mobile device. However, along with the massive computations required to render real-time 3D video games, other significant challenges arises, such as avoiding response delays, and dealing with high bandwidth transmission while ensuring users' Quality of Experience (QoE). For example, ``first person shooter" games have very short delay tolerances (about 100ms \cite{shea2013cloud}). 
If the latency increases, players may leave a game because of worsened interactive experiences. When there are large numbers of simultaneous users, bandwidth requirements may explode. When transmitting over traffic-stressed networks, unstabilities and errors can greatly degrade users' QoE.

Being able to monitor distorted video quality on the client side, sending feedback to cloud servers to adjust their encoding recipes when streaming video can decrease computation and bandwidth requirements while enabling high QoE at the users' side. The basic tools are objective video quality assessment (VQA) algorithms, of which several are currently utilized in streaming and social video applications over very large scales, to produce perceptually accurate predictions with low computational expense.

In the cloud gaming space, reference videos are not accessible on the client side; hence NR VQA models provide reasonable solutions. Many general-purpose NR VQA models have been designed to predict the perceived qualities of real-world video types. Early models, operated by extracting features defined under spatial natural video statistics models, include NIQE \cite{mittal2012making}, BRISQUE \cite{mittal2012no}, HIGRADE \cite{kundu2017no}, GM-LOG \cite{xue2014blind}, and FRIQUEE \cite{ghadiyaram2017perceptual}, among others. Since analyzing temporal distortions is essential to the prediction of dynamic video quality, space-time VQA models have been devised, including V-BLIINDS \cite{saad2014blind}, ChipQA \cite{ebenezer2020no, ebenezer2021chipqa} and RAPIQUE \cite{tu2021rapique}, which computes spatial and temporal bandpass statistical features, along with semantic features computed by a Convolutional Neural Network (CNN). However, gaming videos rendered by computer graphics engines differ statistically from naturalistic videos, which effects the relevance and performances of existing NR VQA models. Indeed, existing methods struggle on recent gaming video databases \cite{barman2018evaluation,barman2020objective,zadtootaghaj2020quality}.

\begin{figure*}[!ht]
    \centering
	\centerline{\includegraphics[scale=0.23]{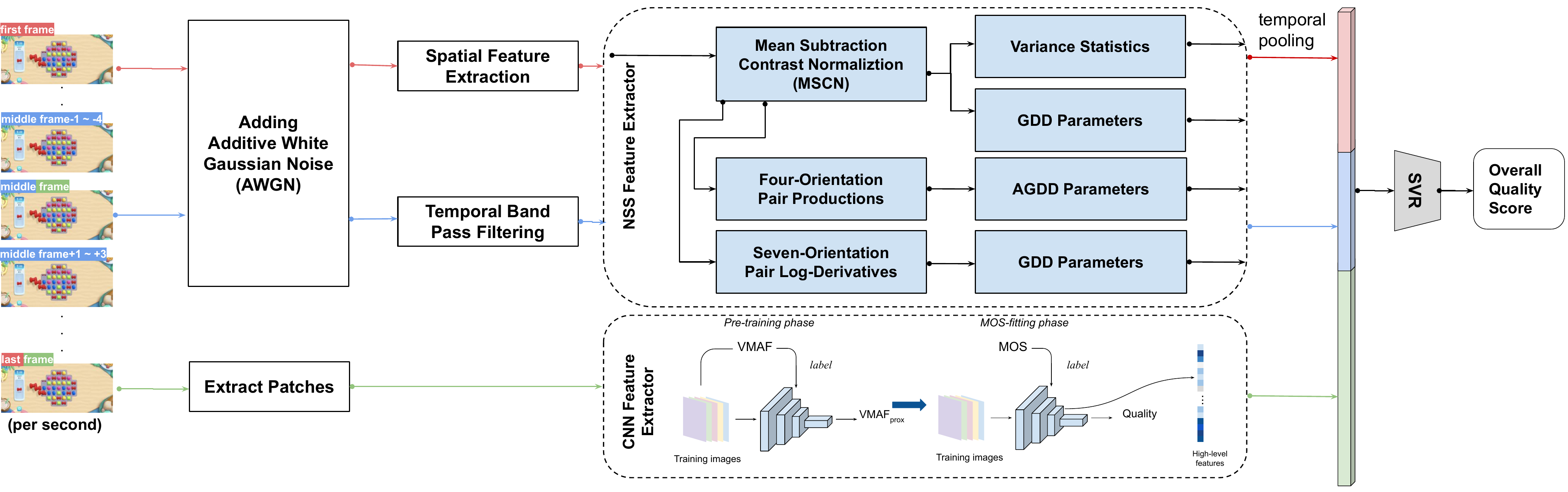}}
    \caption{Schematic flow diagram of the GAMIVAL model. The top portion depicts the spatial and temporal NSS feature computations. The lower portion shows the CNN feature extraction process following NDNetGaming \cite{utke2020ndnetgaming}. All of the features are concatenated and utilized to train an SVR model.}
    \label{fig:model_overview}
\end{figure*}

In recent years, the success of CNNs on many image analysis problems \cite{simonyan2014very, krizhevsky2012imagenet, girshick2014rich}, has motivated their application to NR VQA \cite{li2019quality, ying2021patch}. CNN based VQA models have been shown to perform well on existing UGC datasets, but training them requires adequately large numbers of labeled image samples. Since the advent of widespread cloud gaming technologies and live streaming gaming services over the past decade, only six gaming VQA databases \cite{barman2018gamingvideoset,barman2019no,zadtootaghaj2020quality,wen2021subjective,yu2022subjective,avinab2022mcg} have been proposed. However, unlike large-scale UGC VQA databases containing thousands of videos and highly diverse content, existing gaming video databases contain only a few labeled videos and fewer unique contents. Toward addressing these data limitations, several previous NR gaming VQA models \cite{barman2019no, zadtootaghaj2018nr, goring2019nofu} have been trained and evaluated using labels generated by the full-reference quality metric, Video Multi-Method Assessment Fusion (VMAF) \cite{li2016toward}, which has been shown to deliver superior performance on gaming content datasets \cite{barman2018evaluation}. Using VMAF to generate quality labels makes it possible to train deep networks from scratch, although VMAF is not a perfect predictor of perceptual quality. Another approach that is often effective is transfer learning. The aim is to transfer knowledge learned from one or more source tasks, even if the training and test datasets have somewhat different data distributions \cite{pan2009survey}, which can be rectified by fine-tuning on the target data. For example, a CNN-based NR VQA model NDNetGaming \cite{utke2020ndnetgaming} was first trained on 243,000 images with associated VMAF scores serving as proxy perceptual quality labels, and then fine-tuned on a smaller dataset. This approach was able to achieve high correlations against the subjective scores for gaming content \cite{barman2018gamingvideoset,barman2019no}.

Here we design an NR VQA model called GAMIVAL, which combines the merits of conventional feature-based VQA algorithms with deep CNN-based VQA models. It achieves superior performance with low computational complexity on LIVE-Meta MCG dataset \cite{avinab2022mcg}, as compared to state-of-the-art NR VQA algorithms.

\section{Gaming Video Quality Evaluator (GAMIVAL)}

Fig. \ref{fig:model_overview} shows the processing flow of the GAMIVAL VQA model. It employs the spatial and temporal components of the RAPIQUE model \cite{tu2021rapique}, and the CNN-based features from NDNetGaming \cite{utke2020ndnetgaming}. It also employs a simple ``neural noise" model. These features are concatenated and further used to train an SVR model.

\subsection{Spatial Domain Features + Neural Noise}

Previous spatial-temporal bandpass statistics-based video quality models, like RAPIQUE \cite{tu2021rapique}, have been shown to efficiently capture the perceptual impacts of complex real-world distortions. In these models, bandpass and divisive normalization processes yield mean-subtracted, contrast-normalized (MSCN) coefficients applied on the input image (or on a previously computed feature map) $I(i,j)$:

\begin{equation}
    \hat{I} = \frac{I(i,j) - \mu(i,j)}{\sigma(i,j)+C},
    \label{equ:MSCN}
\end{equation}
where $(i,j)$ are spatial indices, $C$ = 1 is a saturation constant that prevents instabilities, and $\mu$ and $\sigma$ are weighted local means and standard deviations \cite{mittal2012making,mittal2012no} within a gaussian-weighted spatial window centered at location $(i,j)$.

\begin{figure}[!t]
	\centering
	\footnotesize
	\renewcommand{\tabcolsep}{1.3pt} 
	\renewcommand{\arraystretch}{1.3} 
	\begin{tabular}{c cc cc}
        \includegraphics[height=50pt]{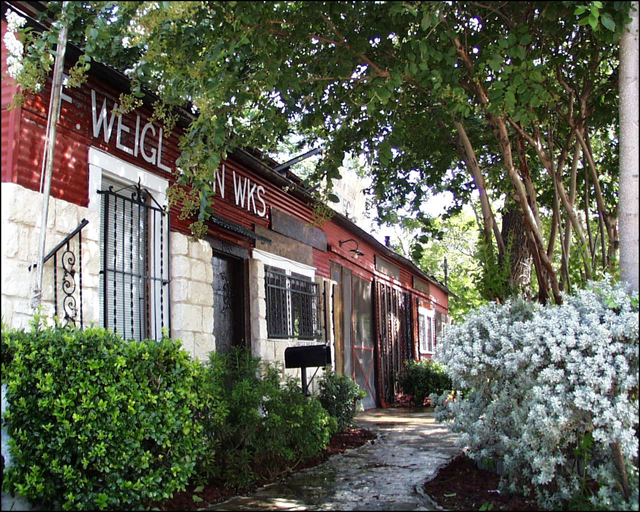} & &
        \includegraphics[height=50pt]{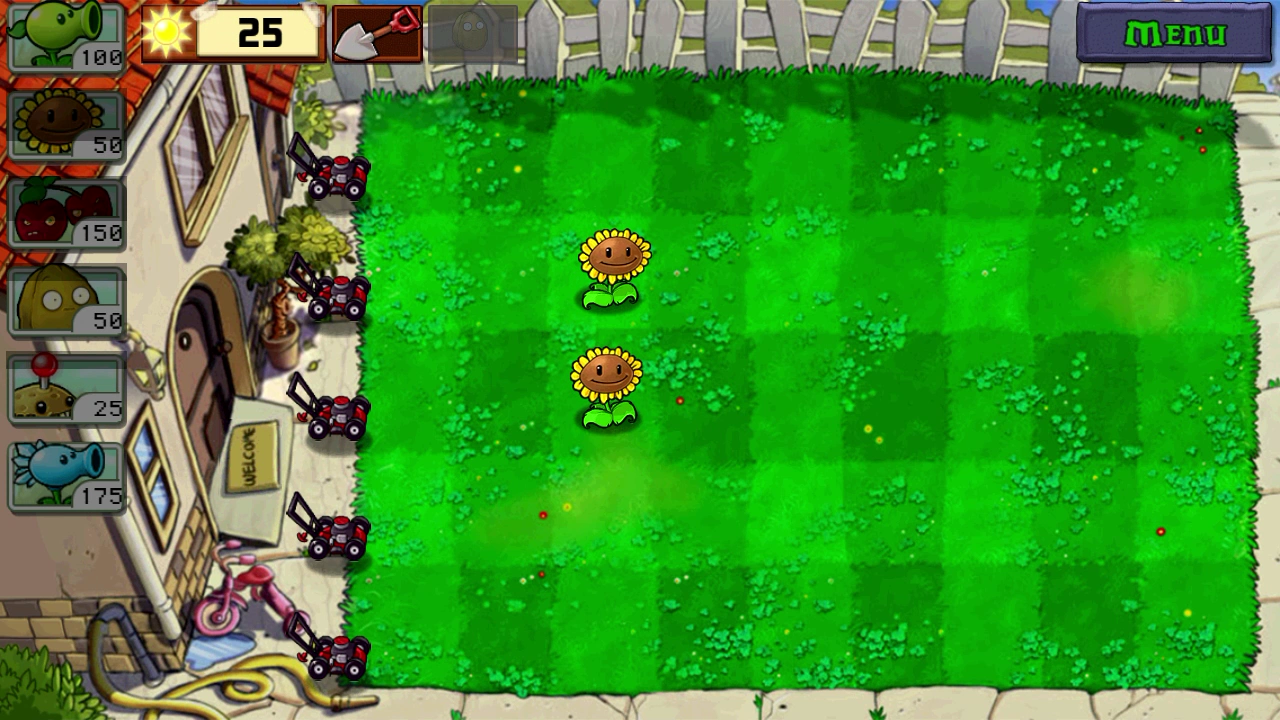} & &
        \includegraphics[height=50pt]{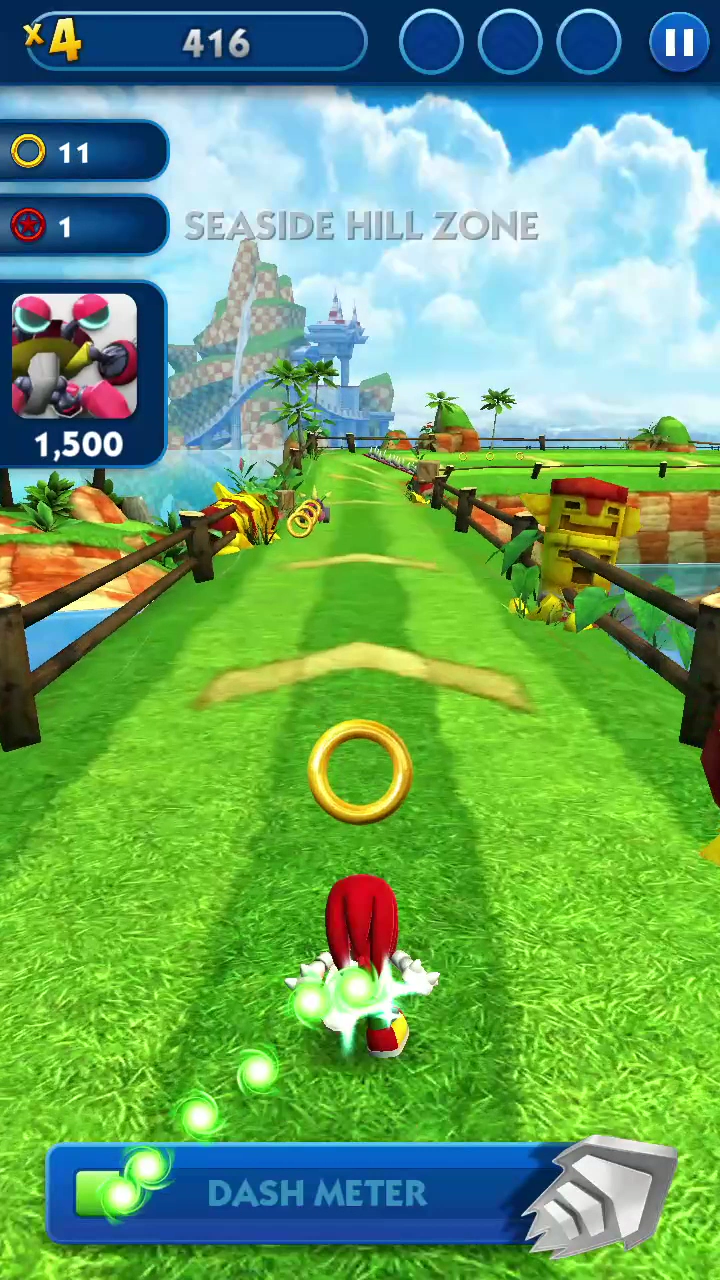}\\
        (a) building2 & & (b) Plants vs. Zombies & & (c) Sonic\\
	\end{tabular}
   \caption{Exemplar test images and frames: (a) natural image from the LIVE IQA database \cite{sheikh2006statistical}, and (b)-(c) two gaming video frames from the LIVE-Meta MCG database \cite{avinab2022mcg}.}
	\label{fig:exemplar}
\end{figure}
It has been widely observed that the bandpass MSCN coefficients of natural image or video frames reveal an underlying statistical regularity. However, visual contents rendered by computer graphics, like gaming videos, typically contain fewer details, and are generally smoother, hence their bandpass statistics differ from those of naturalistic videos or images. Feature computations on these regions can be less stable. Following \cite{jin2021foveated}, we introduce a neural noise model:

\begin{equation}
    \widetilde{I}(i,j) = I(i,j) + W_s
\end{equation}
by which we add white Gaussian noise $W_s \sim N(0,\sigma^2_{W_s})$ to the image before computing the MSCN coefficients.

\begin{figure}[!t]
	\centering
	\footnotesize
	\renewcommand{\tabcolsep}{1.3pt} 
	\renewcommand{\arraystretch}{1.3} 
	\begin{tabular}{c cc}
        \includegraphics[height=80pt]{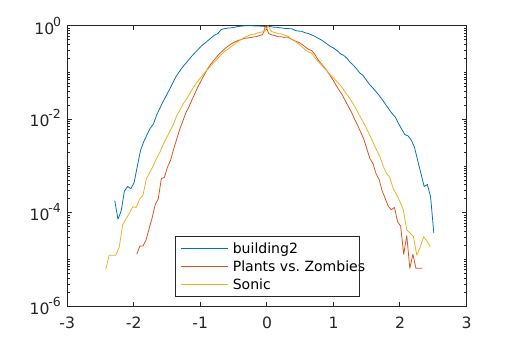} & &
        \includegraphics[height=80pt]{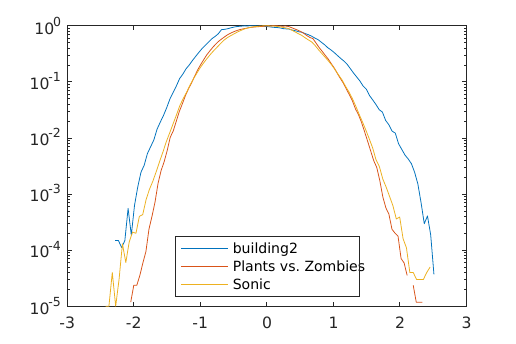} \\
        \includegraphics[height=80pt]{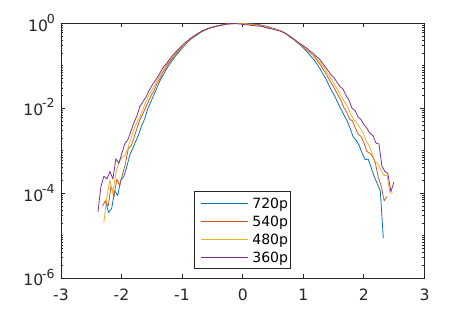} & &
        \includegraphics[height=80pt]{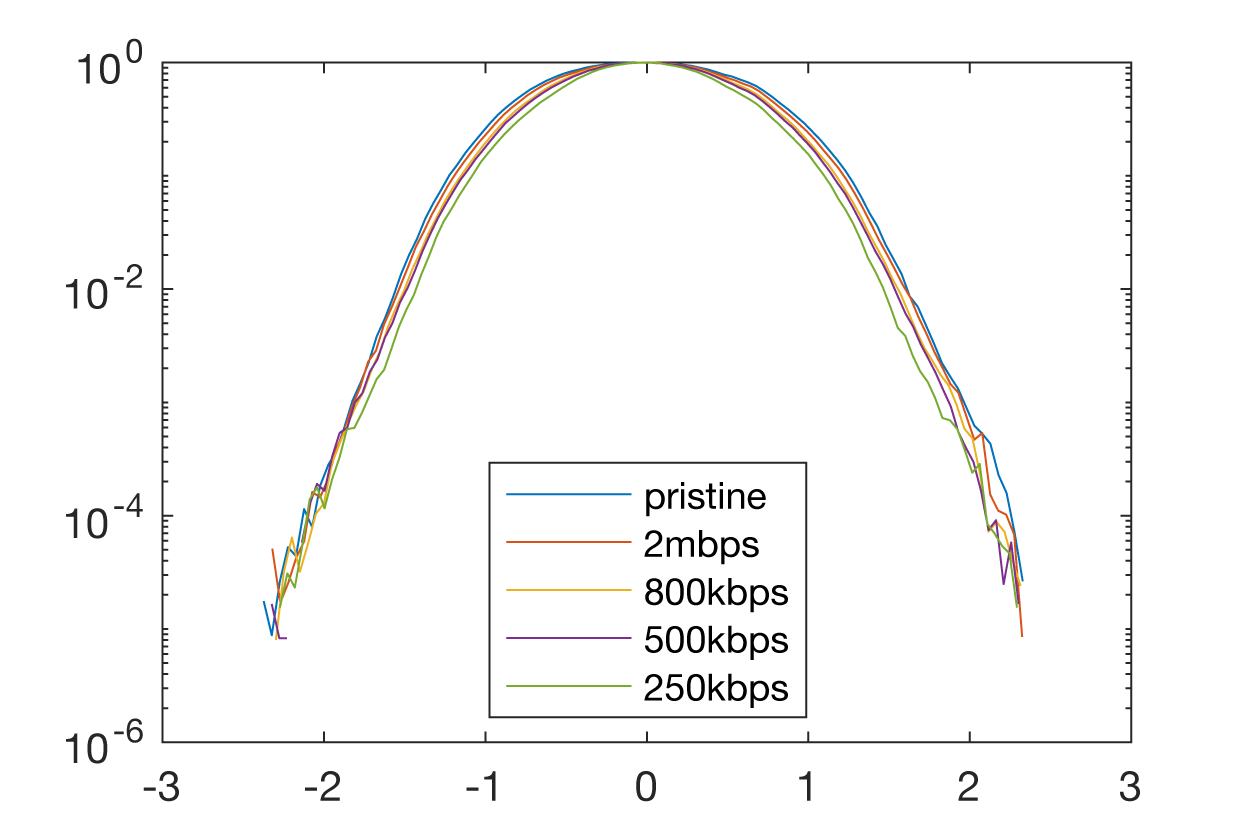} \\
    \end{tabular}
   \caption{Histograms of the MSCN coefficients of the one natural image and two gaming video frames shown in Fig. \ref{fig:exemplar} before adding noise (upper left) and after adding noise (upper right), and the MSCN coefficients of the gaming video ``State of Survival" over varying resolutions (lower left) and bitrates (at 720p) (lower right).}
	\label{fig:noisy}
	\label{fig:compression}
\end{figure}

As in \cite{jin2021foveated}, we have observed that the distributions of the MSCN coefficients of high quality gaming content video frames tends towards Gaussianity. To visualize this, we selected one natural image from the LIVE IQA database \cite{sheikh2006statistical}: building2, and two pristine video frames (720p) from two games in the LIVE-Meta MCG database \cite{avinab2022mcg}: Plants vs. Zombies and Sonic, as shown in Fig. \ref{fig:exemplar}. Fig. \ref{fig:noisy} top portion shows the histograms of the MSCN coefficients of each of these images before and after adding the simulated neural noise ($\sigma_{W_s}$=1.5). It may be observed that histograms of the gaming video frames contain singular spikes, but after the noise was added to the gaming video frames, their MSCN histograms became very similar to those of the natural image, with a Gaussian appearance. Another gaming frame (State of Survival) was chosen to further visualize how these regularities are affected by distortion. Fig. \ref{fig:compression} lower portion shows the MSCN coefficients of this frames after resampling to different resolutions and bitrates. Quantifying these deviations to predict perceived quality is effectively accomplished using the basic statistical feature extraction module NSS-34 in RAPIQUE \cite{tu2021rapique}. We apply this module to the luma and chromatic feature maps at two scales (the original scale resized to 540p and 270p) as in RAPIQUE \cite{tu2021rapique}, by uniformly sampling 2 frames per second, thus producing 680 spatial features.

\subsection{Temporal Domain Features + Neural Noise}

\begin{figure}[!t]
	\centering
	\footnotesize
	\renewcommand{\tabcolsep}{1.3pt} 
	\renewcommand{\arraystretch}{1.3} 
	\begin{tabular}{c cc}
        \includegraphics[height=75pt]{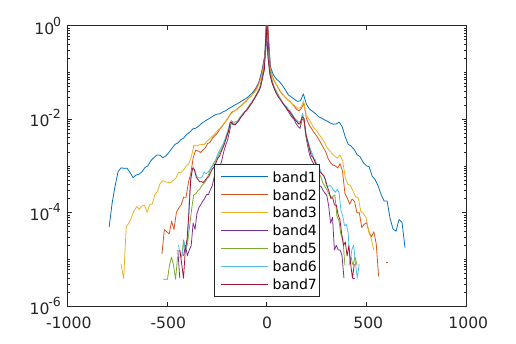} & &
        \includegraphics[height=75pt]{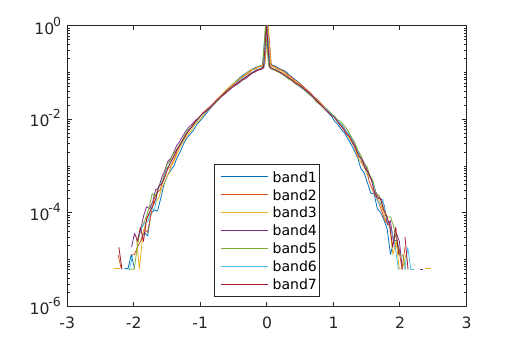}\\
        \includegraphics[height=75pt]{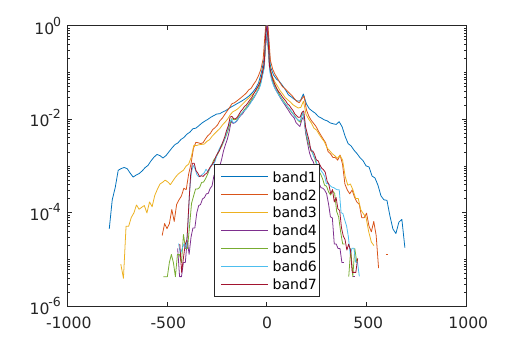} & &
        \includegraphics[height=75pt]{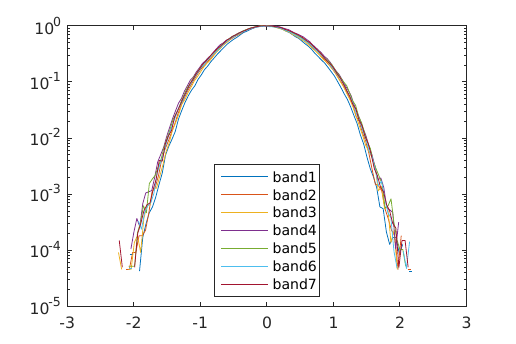}\\
    \end{tabular}
   \caption{Histograms of raw temporal subband coefficients \textbf{before} adding noise (upper left) and \textbf{after} adding noise (lower left), and the corresponding spatial MSCN coefficients of the gaming video ``PGA Golf Tour" \textbf{before} adding noise (upper right) and \textbf{after} adding noise (lower right).}
	\label{fig:temporal}
	\label{fig:temporal_noisy}
\end{figure}

Since temporal bandpass statistics have been shown to be predictive of frame rate-dependent video quality \cite{madhusudana2021st}, the authors of RAPIQUE \cite{tu2021rapique} proposed a temporal model utilizing temporal bandpass coefficients and subsequentially applying spatial MSCN transforms, as in Eq. (\ref{equ:MSCN}). The MSCN coefficients of temporal bandpass coefficients of gaming videos also exhibit discontinuous histograms, as shown in Fig. \ref{fig:temporal} top portion. Hence again apply additive neural noise before computing the spatial MSCN coefficients

\begin{equation}
    \widetilde{Y_k}(\textbf{x},t) = {Y_k}(\textbf{x},t) + W_t
\end{equation}
as before, where ${Y_k}(x,t)$ are the temporal bandpass coefficients, $k=1,...,7$ denotes subband indices, $\textbf{x}=(x,y)$ and \emph{t} are spatial and temporal coordinates, and $W_t\sim N(0,\sigma^2_{W_t})$ is the noise added to the temporal model.

As shown in Fig. \ref{fig:temporal_noisy} lower portion, after adding the noise ($\sigma_{W_t}$=1.5), the histograms of the MSCN coefficients of temporal bandpass coefficients of gaming videos also present Gaussian appearances. These regularities are modified by the presence of distortions, which provides a way of quantifying deviations, and therefore, to predict video quality scores. Towards this end, we deploy the NSS-34 operator set in RAPIQUE \cite{tu2021rapique} on the temporal bandpass coefficients of each analyzed gaming video at two scales (the original scale resized to 540p and 270p), sampling at 8 frames per second and then applying a temporal Haar filter as in \cite{tu2021rapique} to extract 7 bandpass responses, yielding 34$\times$7$\times$2 = 476 temporal features at each time sample. Section \ref{appendix:A} provides additional information on the selection of the parameters $\sigma_{W_s}$ and $\sigma_{W_t}$ of the additive noise elements ${W_s}$ and ${W_t}$ in equations (2) and (3), respectively. It is also shown that model performance significantly improves by the addition of noise, and remains robust over even orders of magnitude of the parameters.

\subsection{CNN-based Features}
Since existing gaming video content datasets contain too few videos to allow training of CNN feature extractors from scratch, we instead applied NDNetGaming model \cite{utke2020ndnetgaming}. Unlike the CNN branch in RAPIQUE \cite{tu2021rapique}, (a pre-trained ResNet-50 model \cite{he2016deep}), the authors of \cite{utke2020ndnetgaming} instead fine-tuned a DenseNet-121 model \cite{huang2017densely}) in two steps: first, they retrained the last 57 convolutional layers of a DenseNet-121 by replacing the fully connected (FC) layer with a dense layer, using VMAF values as proxy subjective quality labels. Second, they fine-tuned the resulting pre-trained model by retraining the last 36 layers based on human subjective image quality ratings. Finally, they assigned a different weight to each frame, based on its temporal complexity. Without being retrained on different cloud game video databases, this model might not deliver robust and generalized performance. The authors of \cite{tu2021ugc} showed that, without fine-tuning, the simple feature vector of an FC layer could be a useful quality indicator if a shallow regressor is trained on top. Therefore, we discarded the pooling layer of the original NDNetGaming model, thereby yielding 1024 activation features. By also taking into account the time complexity, we also devised a temporal sampling strategy. Instead of extracting features on every frame, the CNN backbone (NDNetGaming) operates at 2 frame per second.

\subsection{Quality Evaluation}
After obtaining all of the spatial, temporal, and CNN-based features within each one-second chunk, they are concatenated into a 2180-dimensional feature vector. By average-pooling the vectors within each video chunk, a vector of features is obtained for the entire video and then trained with a shallow or deep regressor head. In our implementation, we used a support vector machine regressor (SVR) to map the features to predicted video quality scores \cite{mittal2012no,saad2014blind,ghadiyaram2017perceptual,korhonen2019two}.
\begin{figure}[!t]
	\centering
	\footnotesize
	\renewcommand{\tabcolsep}{1.3pt} 
	\renewcommand{\arraystretch}{1.3} 
	\includegraphics[height=75pt]{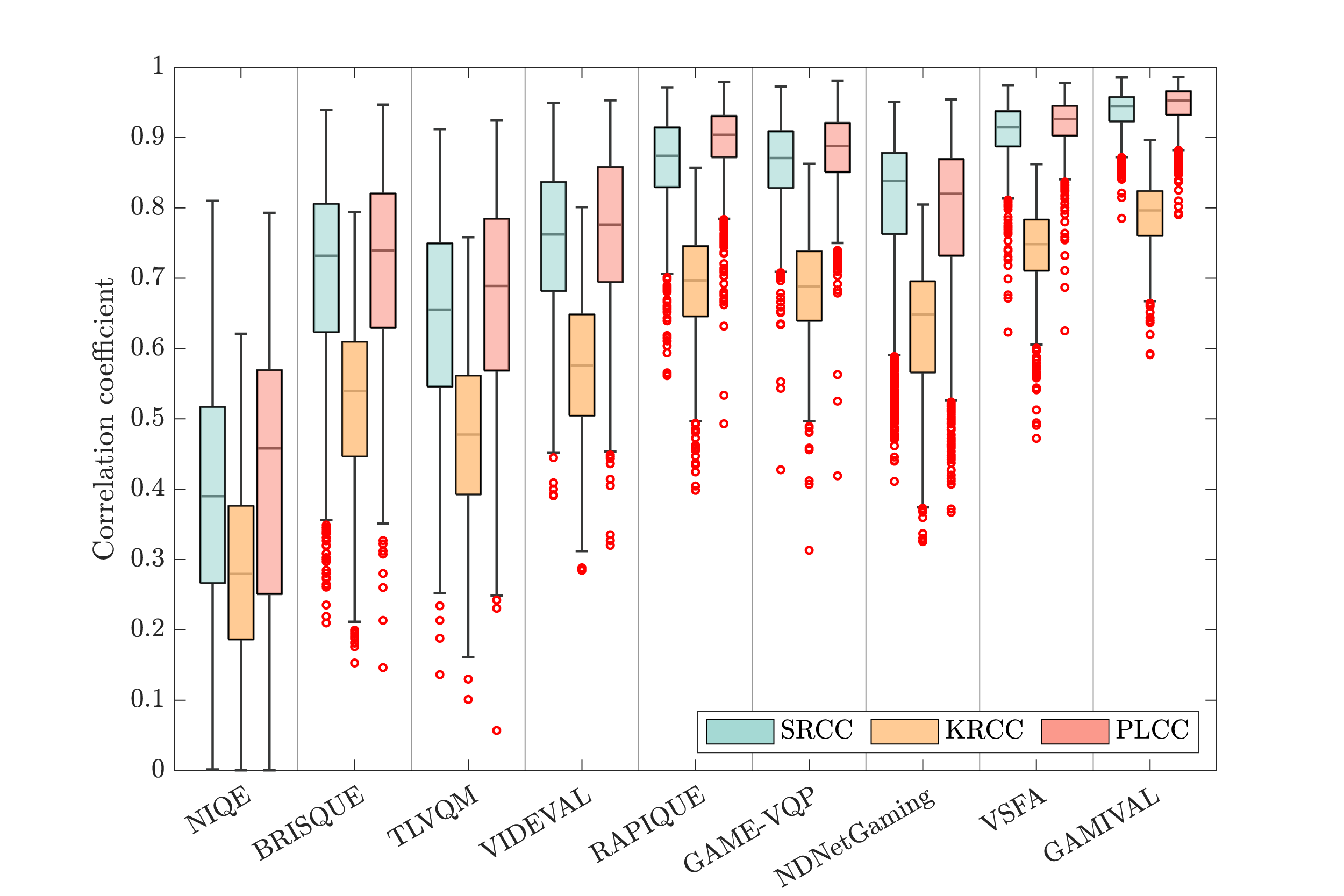}
   \caption{Box plots of PLCC, SRCC, and KRCC of evaluated BVQA algorithms on the LIVE-Meta MCG dataset over 1000 splits. For each box, the median is indicated by the center line, while the box edges represent the 25th and 75th percentiles, and outliers are indicated by red circles.}
	\label{fig:boxplot}
\end{figure}

\section{Performance evaluation}

\subsection{The LIVE-Meta MCG Database}
To evaluate the performance of our model against other methods, we used the recently created LIVE-Meta MCG database \cite{avinab2022mcg}. This database contains mean opinion scores (MOS) of 600 landscape and portrait gaming videos generated from 30 pristine source sequences obtained from 16 different games using 20 different resolution-bitrate pairs to compress each pristine video. The resolution ranged from 360p to 720p, while the bitrates range from 250 kbps to 2 mbps.

The subjective human study was conducted on a Google Pixel 5 mobile device. All of the videos were upscaled to fit the mobile screen size (1080p) using FFMPEG's default bicubic interpolation function following decoding to native videos. To map features to the MOS that was obtained from users viewing a 1080p Pixel 5 display, we applied all of the compared algorithms on the display resolution.

\subsection{Evaluation Framework}
\subsubsection{Compared Methods}
We evaluated the performances of several popular NR IQA and VQA models: NIQE \cite{mittal2012making}, BRISQUE \cite{mittal2012no}, TLVQM \cite{korhonen2019two}, VIDEVAL \cite{tu2021ugc}, RAPIQUE \cite{tu2021rapique}, VSFA \cite{li2019quality}, and two gaming VQA models: NDNet-Gaming \cite{utke2020ndnetgaming} and GAME-VQP \cite{yu2022perceptual}. NIQE is an unsupervised model that pools predicted frame quality scores to generate video quality predictions. Supervised VQA methods like BRISQUE, TLVQM, VIDEVAL, RAPIQUE, and GAME-VQP operate by training an SVR to learn feature-to-score mappings. The deep learning-based model VSFA uses a Resnet-50 \cite{he2016deep} CNN backbone to obtain quality-aware features, then maps them to MOS using a single FC layer and a Gated Rectified Unit (GRU). NDNet-Gaming regresses video quality predictions using a DenseNet-121 \cite{huang2017densely} backbone.

\begin{table}[!t]
   \renewcommand{\arraystretch}{1.3}
   \caption{Median SRCC, KRCC, PLCC and RMSE on the LIVE-Meta MCG database over 1000 train-test splits. The \underline{underlined} and \textbf{boldfaced} entries represent the best and top three performers.}
   \label{tab:comparison}
   \tiny
   \setlength{\tabcolsep}{3pt}
   \centering
   \renewcommand{\tabcolsep}{3.3pt} 
   \begin{tabular}{rcccc}
   \toprule[1.2pt]
   Metrics & SRCC($\uparrow$) & KRCC($\uparrow$) & PLCC($\uparrow$) & RMSE($\downarrow$) \\
   \midrule[0.5pt]
   NIQE~  & ~-0.3900~ & ~-0.2795~ & ~0.4581~  & ~16.5475~ \\
   BRISQUE~  & ~0.7319~ & ~0.5395~ & ~0.7394~  & ~12.5618~ \\
   TLVQM~    & ~0.6553~ & ~0.4777~ & ~0.6889~ & ~13.5413~ \\
   VIDEVAL~    & ~0.7621~ & ~0.5756~ & ~0.7763~  & ~11.7520~ \\
   RAPIQUE~    & ~\textbf{0.8740}~ & ~\textbf{0.6964}~ & ~\textbf{0.9039}~  & ~\textbf{8.0242}~ \\
   GAME-VQP~    & ~0.8709~ & ~0.6885~ & ~0.8882~  & ~8.5960~ \\
   NDNet-Gaming~    & ~0.8382~ & ~0.6485~ & ~0.8200~  & ~10.5757~ \\
   VSFA~    & ~\textbf{0.9143}~ & ~\textbf{0.7484}~ & ~\textbf{0.9264}~  & ~\textbf{7.1316}~ \\
   \midrule[0.5pt]
   GAMIVAL~    & ~\underline{\textbf{0.9441}}~ & ~\underline{\textbf{0.7963}}~ & ~\underline{\textbf{0.9524}}~  & ~\underline{\textbf{5.7683}}~ \\
   \bottomrule[1.2pt]
\end{tabular}
\end{table}

\subsubsection{Evaluation Method}
We randomly split the dataset into training and test sets (80\%/20\%), by content, over 1000 iterations. The training set was further split to conduct five-fold cross-validation. When splitting the training and validation sets, we also ensured that the contents were mutually disjoint. We optimized the SVR parameters $(C,\gamma)$ using grid search on the training and validation sets. All of the evaluated supervised VQA methods were trained and tested using the previous mentioned split strategy. NIQE is an unsupervised model, hence was not trained. NDNet-Gaming is based on a pretrained model and was evaluated on 1000 test splits without training. While testing VSFA, the training and validation sets were used to optimize the best-performing FC-GRU model weights. Four performance metrics were used to evaluate algorithm performance: the Spearman's Rank-Order Correlation Coefficient (SRCC), the Kendall Rank Correlation Coefficient (KRCC), Pearson's Linear Correlation Coefficient (PLCC), and the Root Mean Square Error (RMSE). The median values of these four metrics against MOS are reported.

\subsection{Evaluation Results of NR-IQA and VQA Models}

Table \ref{tab:comparison} shows the model performances on the LIVE-Meta MCG database. It may be observerd that GAMIVAL achieved the best performance, while VSFA and RAPIQUE ranked second and third respectively. It is worth noting that the two gaming VQA models (NDNet-Gaming and GAME-VQP) also yielding fairly good fair performance. For better visualization, Fig. \ref{fig:boxplot} shows box plots of the SRCC, KRCC, and PLCC values. It may be noticed that the GAMIVAL values are more tightly grouped, indicating its stable performance and low variance.

Section \ref{appendix:B} includes an ablation study that analyzes the contributions of each GAMIVAL feature set. We further summarize the complexity and runtime comparison of the NR-VQA models in Section \ref{appendix:C}.

\section{Conclusion and future work}

We have developed a dual path gaming video quality assessment model that deploys distortion-sensitive natural scene features along one path, and distortion and semantically aware deep features along the other path. In order to better regularize the distributions of the space-time MSCN video coefficients, especially on extremely smooth or constant regions often found in gaming content videos, we applied an additive ``neural noise" mechanism which led to much improved prediction performance. Evaluations on a recent large-scale MCG video database show that the new model, GAMIVAL, achieves state-of-the-art quality prediction accuracy with low computational complexity as compared with leading conventional and deep learning based VQA models.

\clearpage

\section{Appendix}

\begin{table}[!t]
   \renewcommand{\arraystretch}{1.3}
   \caption{Median SRCC, KRCC, PLCC and RMSE on the LIVE-Meta MCG database over 100 train-test splits.}
   \label{tab:noise}
   \scriptsize
   \setlength{\tabcolsep}{3pt}
   \centering
   \renewcommand{\tabcolsep}{3.3pt} 
   \begin{tabular}{rcccc}
   \toprule[1.2pt]
   $\sigma_{W_s}, \sigma_{W_t}$ & SRCC($\uparrow$) & KRCC($\uparrow$) & PLCC($\uparrow$) & RMSE($\downarrow$) \\
   \midrule[0.5pt]
   0~  & ~0.8949~ & ~0.7252~ & ~0.9108~  & ~7.5875~ \\
   0.01~  & ~0.9371~ & ~0.7878~ & ~0.9496~  & ~5.9064~ \\
   0.05~  & ~0.9376~ & ~0.7878~ & ~0.9490~  & ~5.8108~ \\
   0.1~    & ~0.9374~ & ~0.7887~ & ~0.9520~ & ~\textbf{5.6773}~ \\
   0.3~    & ~0.9407~ & ~0.7893~ & ~0.9521~  & ~\textbf{\underline{5.6177}}~ \\
   0.5~    & ~\textbf{0.9427}~ & ~\textbf{0.7955}~ & ~\textbf{\underline{0.9550}}~  & ~5.7544~ \\
   1~    & ~0.9392~ & ~0.7886~ & ~0.9505~  & ~5.7000~ \\
   1.5~    & ~\textbf{\underline{0.9439}}~ & ~\textbf{\underline{0.7962}}~ & ~\textbf{0.9526}~  & ~5.6941~ \\
   2~    & ~0.9366~ & ~{0.7857}~ & ~{0.9493}~  & ~{5.8625}~ \\
   3~    & ~0.9387~ & ~{0.7893}~ & ~{0.9494}~  & ~{5.7856}~ \\
   \bottomrule[1.2pt]
\end{tabular}
\end{table}

\subsection{Neural Noise}
\label{appendix:A}
To study the parameters $\sigma_{W_s}$ and $\sigma_{W_t}$ of the noise elements $W_s$ and $W_t$ in (2) and (3), we evaluted GAMIVAL on the LIVE-Meta MCG Database while varying these parameters. As may be seen from Table \ref{tab:noise}, the performance of GAMIVAL was remarkably robust over a very wide range of the two noise parameters, which we held equal, even over several orders of magnitude. We selected the best-performing values $\sigma_{W_s} = \sigma_{W_t} = 1.5$, but GAMIVAL significantly outperformed all other models for all nonzero parameter values.

\subsection{Ablation Study}
\label{appendix:B}

\begin{figure}[!t]
	\centering
	\normalsize
	\renewcommand{\tabcolsep}{1.3pt} 
	\renewcommand{\arraystretch}{1.3} 
	\includegraphics[height=100pt]{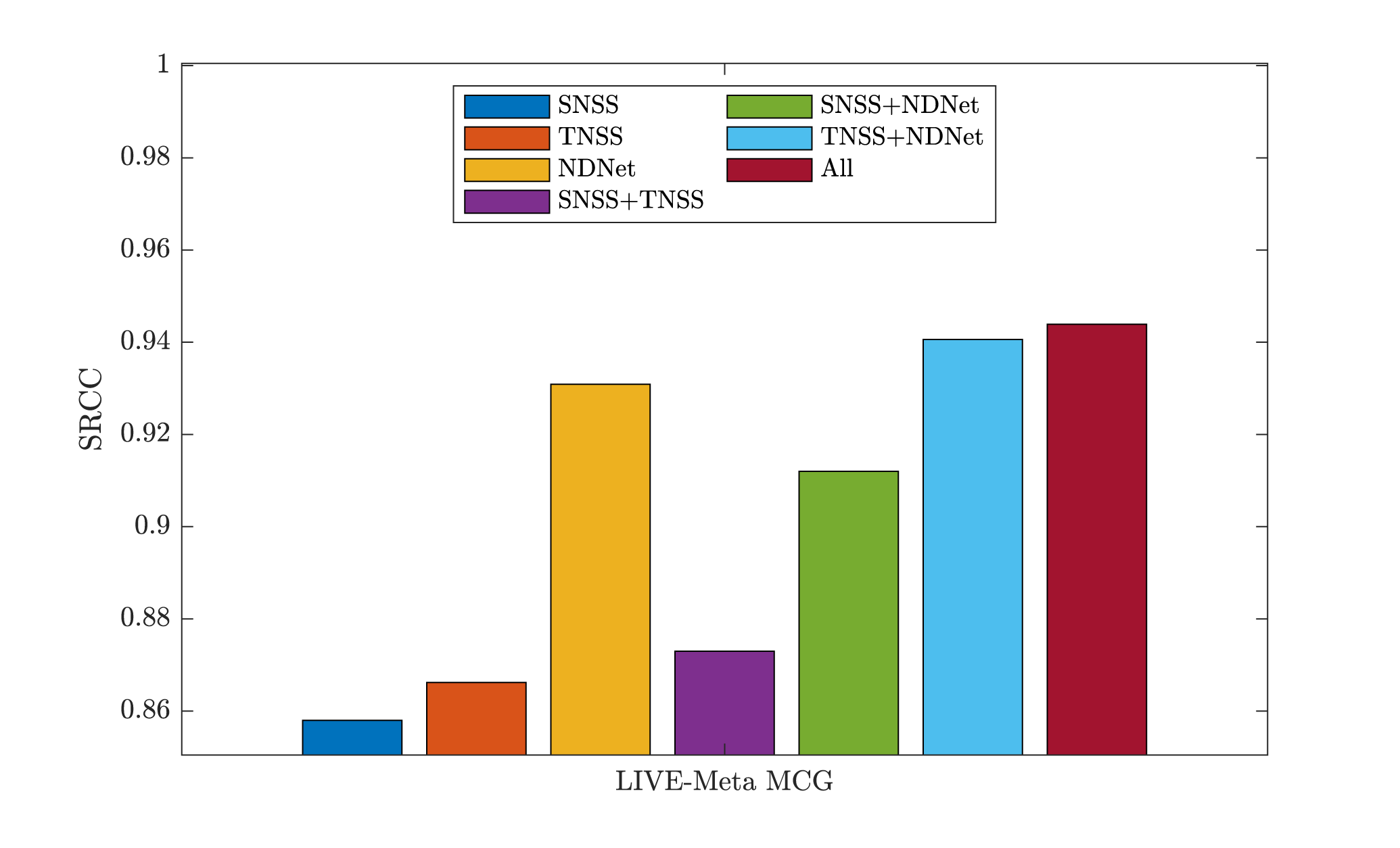}
   \caption{Results of ablation study of GAMIVAL against three feature combinations:Noisy SpatialNSS (SNSS, Sec. II-A), Noisy TemporalNSS (TNSS, Sec. II-B), and (NDNet) deep features (Sec. II-C).}
	\label{fig:ablation}
\end{figure}

To analyze the contribution of each GAMIVAL feature set, we conducted an ablation study. Fig. \ref{fig:ablation} shows the increase in performance obtained as each feature combination is added. It may be observed that the temporal ``additive noise" features improved performance more than did the spatial ``additive noise" features. The NDNetGaming component contributes significantly to the performance, likely because it adds semantic significance to the quality predictions.

\subsection{Complexity and Runtime Comparison}
\label{appendix:C}

\begin{figure}[!t]
	\centering
	\normalsize
	\renewcommand{\tabcolsep}{1.3pt} 
	\renewcommand{\arraystretch}{1.3} 
	\begin{tabular}{c}
        \includegraphics[scale=0.3]{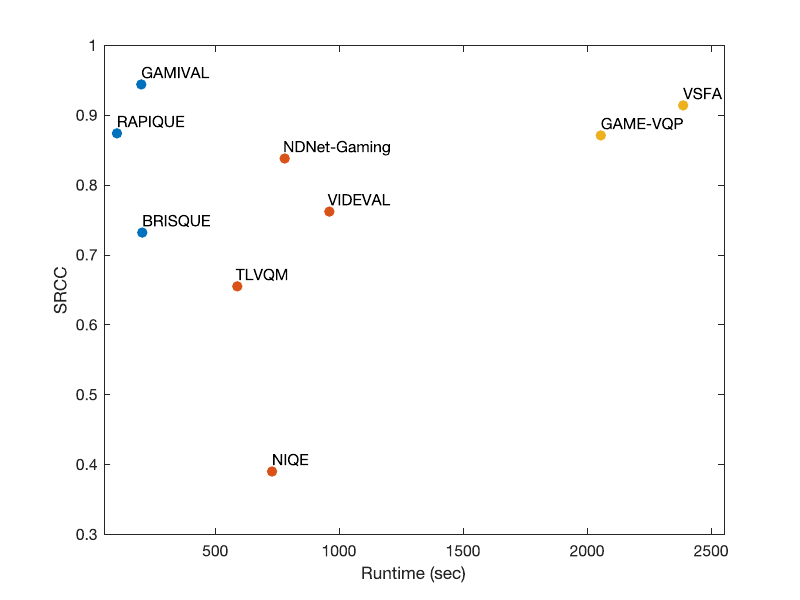}
        \includegraphics[scale=0.3]{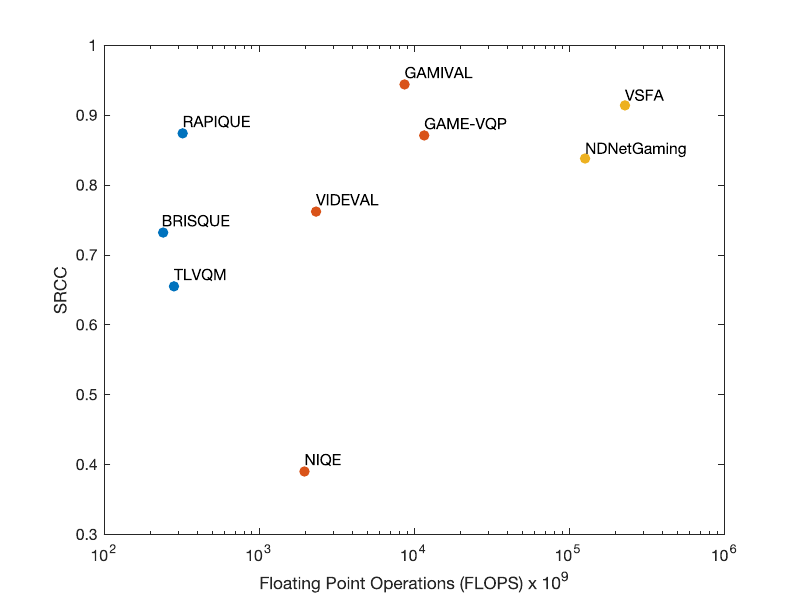}\\
    \end{tabular}
   \caption{Scatter plots of SRCC of NR-VQA algorithms versus runtime on 1080p videos (right) and performance versus FLOPs (left).}
	\label{fig:complexity}
\end{figure}

The experiments were performed in MATLAB R2021a and Python 3.6.10 under Ubuntu 20.04.4 LTS on a Desktop with an Intel Xeon CPU E5-2620 v4@2.10GHz processor, and 64GB RAM. For a fair comparison, all algorithms were carried out on the CPU. We used one 1080x2160 video from the LIVE-Meta MCG database to study the computational efficiency of the NR-VQA models. Table \ref{tab:complexiy} lists the execution time and floating point operations, while Fig. \ref{fig:complexity} shows visual representations of the trade-off between performance and complexity. It may be observed that although VSFA delivers very good performance, it has the highest computational complexity. The top-performing algorithms, RAPIQUE and GAMIVAL, are both computationally efficient, making them promising options for real-time video quality prediction applications.

\begin{figure}[!t]
	\centering
	\footnotesize
	\renewcommand{\tabcolsep}{1.3pt} 
	\renewcommand{\arraystretch}{1.3} 
	\begin{tabular}{c cc cc cc cc}
        \makecell[c]{
        \includegraphics[height=45pt]{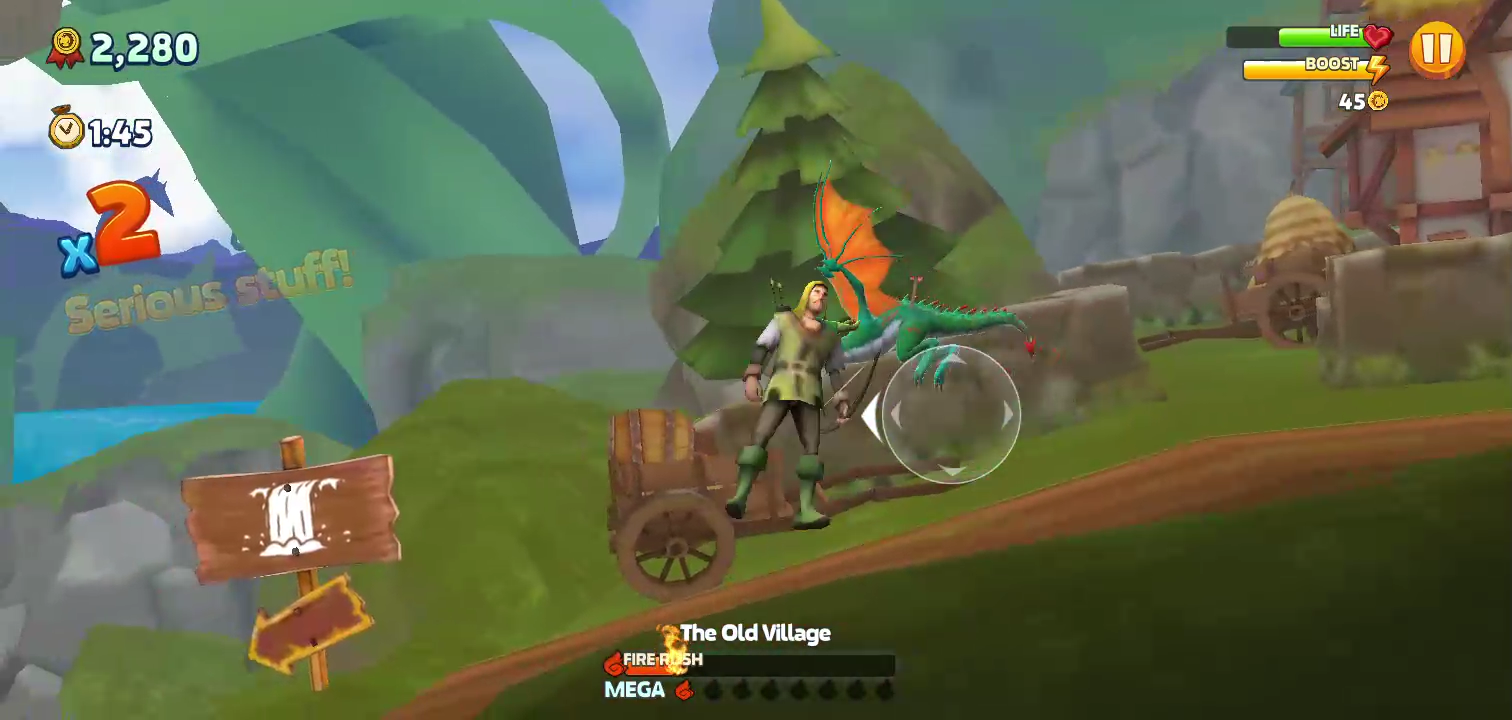} \\
        (a) MOS=73.96,\\
        RAPIQUE=67.14, \\
        GAMIVAL=77.08 \\
        \includegraphics[height=45pt]{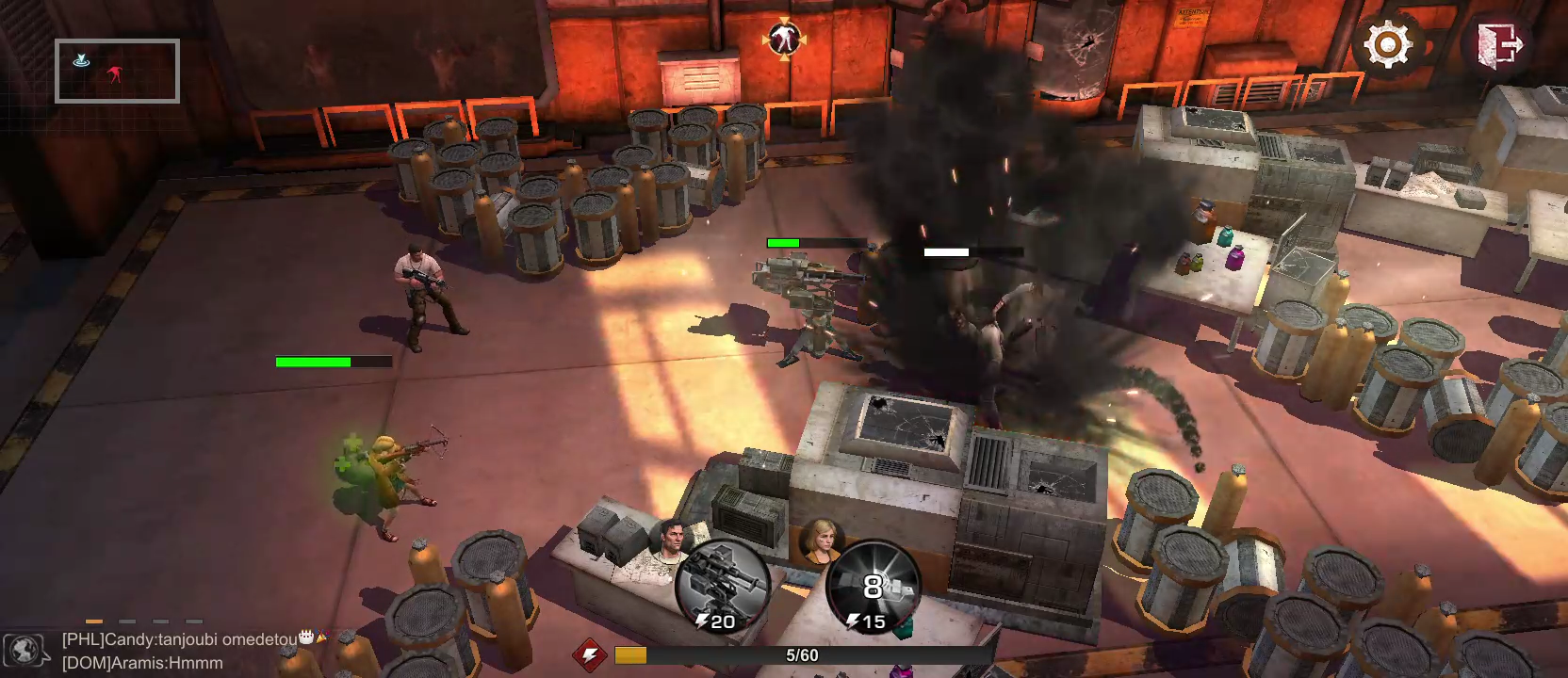}\\
        (b) MOS=73.14,\\
        RAPIQUE=66.38, \\
        GAMIVAL=74.05
        } & &
        & &
        \makecell[c]{
        \includegraphics[height=120pt]{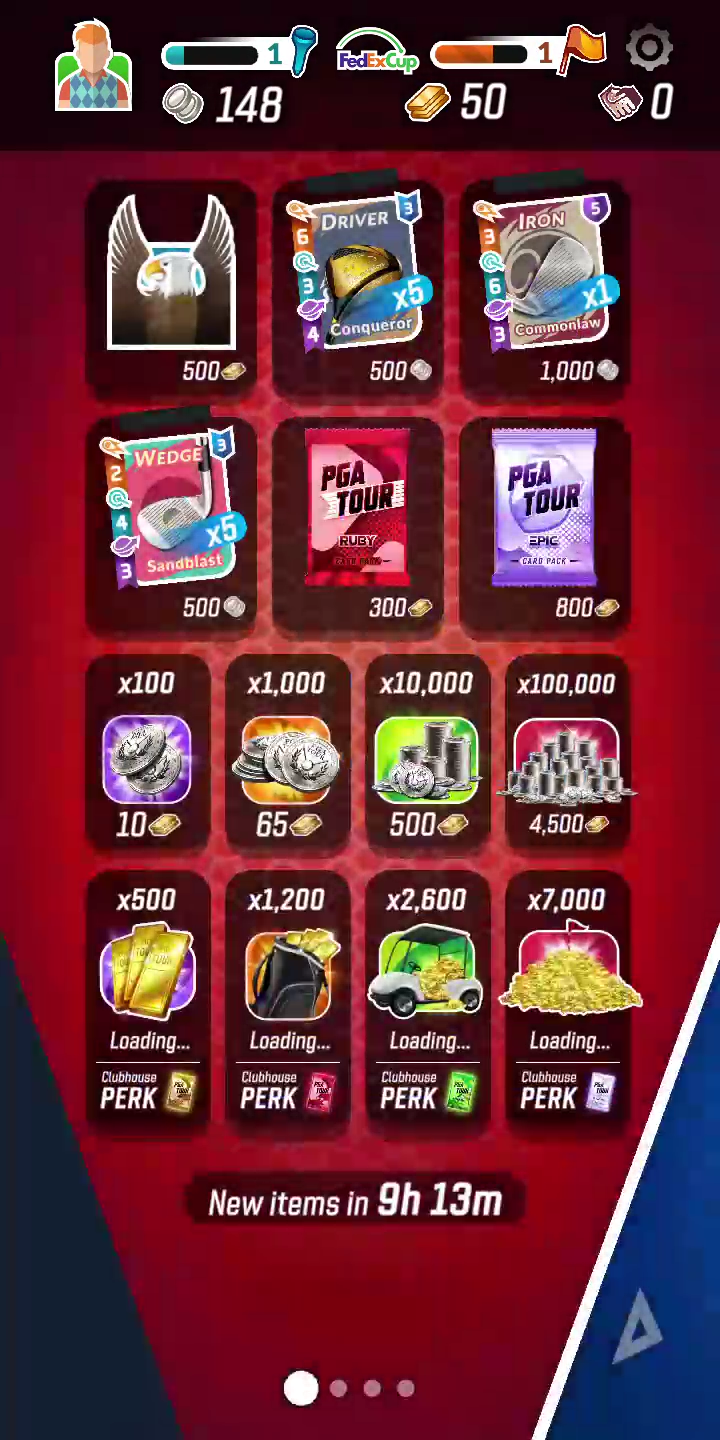} \\
        (c) MOS=51.57,\\
        RAPIQUE=59.55, \\
        GAMIVAL=54.69
        } & &
        & &
        \makecell[c]{
        \includegraphics[height=120pt]{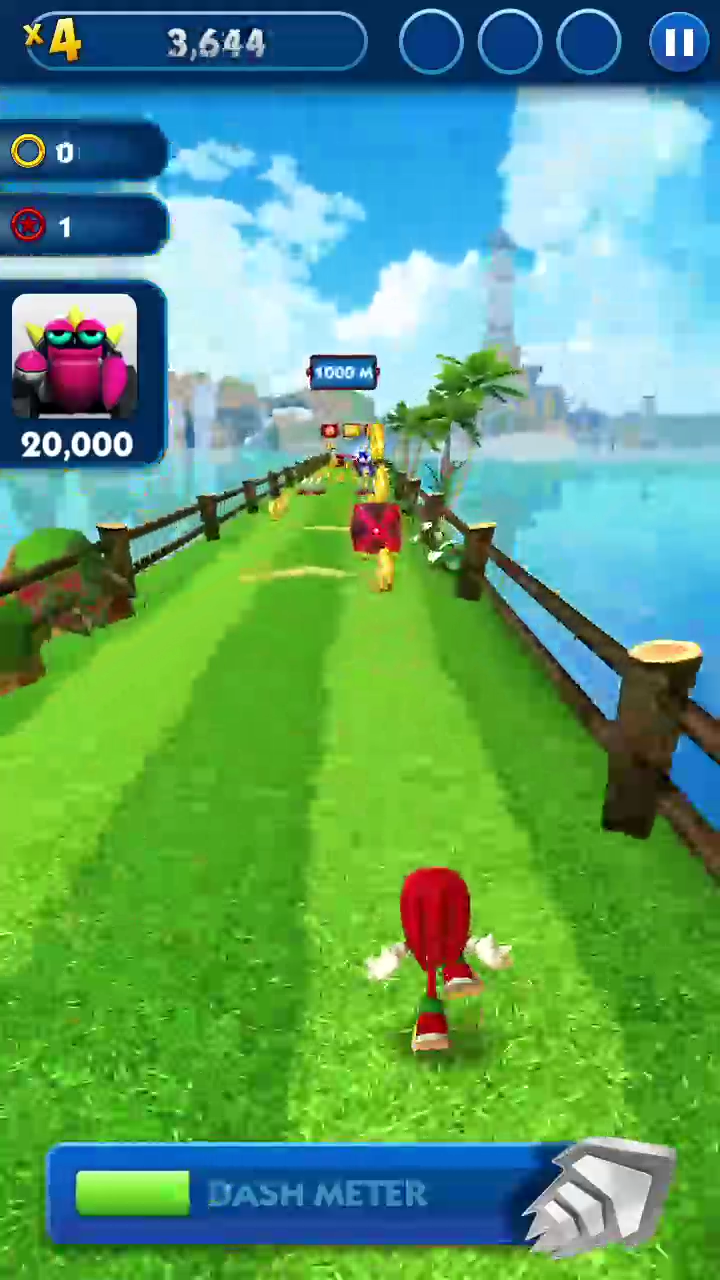} \\
        (d) MOS=40.10,\\
        RAPIQUE=56.78, \\
        GAMIVAL=48.39
        }\\
    \end{tabular}
   \caption{Exemplar test frames of 720p gaming videos having varying bitrate values, taken from the LIVE-Meta MCG Database, along with their MOS and predicted quality score computed by RAPIQUE and GAMIVAL: (a) Hungry Dragon, (b) State of Survival, (c) PGA, (d) Sonic.}
\end{figure}

\begin{table}[!t]
   \renewcommand{\arraystretch}{1.3}
   \caption{Model computation complexity on a 1080p video.}
   \label{tab:complexiy}
   \tiny
   \setlength{\tabcolsep}{3pt}
   \centering
   \renewcommand{\tabcolsep}{3.3pt} 
   \begin{tabular}{rccc}
   \toprule[1.2pt]
   Model & Platform & \makecell[c]{Time\\(seconds)} & \makecell[c]{FLOPS\\($\times10^9$)} \\
   \hline
   NIQE~  & ~MATLAB~ & ~728~ & ~1965~ \\
   BRISQUE~  & ~MATLAB~ & ~\textbf{205}~ & ~\underline{\textbf{241}}~ \\
   TLVQM~    & ~MATLAB~ & ~588~ & ~\textbf{283}~ \\
   VIDEVAL~    & ~MATLAB~ & ~959~ & ~2334~ \\
   RAPIQUE~    & ~MATLAB~ & ~\underline{\textbf{103}}~ & ~\textbf{322}~ \\
   GAME-VQP~    & ~MATLAB~ & ~2053~ & ~11627~ \\
   NDNet-Gaming~    & ~Python, Tensorflow~ & ~779~ & ~126704~ \\
   VSFA~    & ~Python, Pytorch~ & ~2385~ & ~229079~ \\
   \midrule[0.5pt]
   GAMIVAL~    & \makecell[c]{Python, Tensorflow, \\MATLAB}~ & ~\textbf{201}~ & ~8683~ \\
   \bottomrule[1.2pt]
   \end{tabular}
\end{table}
\clearpage
\bibliographystyle{IEEEtran}
\bibliography{IEEEexample}{}
\end{spacing}
\end{document}